\documentstyle[12pt]{article}
\newlength{\mathspace}
\def\ch #1{\check{#1}}
\def\h #1{\hat{#1}}
\def\np#1{ Nucl. Phys. B#1}
\def\pr#1    { Phys. Rev. D#1 }
\def\pl#1{ Phys. Lett. B#1}

\def\ijmp#1  { Int. Jour. Mod. Phys. A#1 }
\def\mpl#1   { Mod. Phys. Lett. A#1 }
\def\begineq{\begin{equation}}
\def\endeq{\end{equation}}
\def\eqabegin{\begin{eqnarray}}
\def\eqaend{\end{eqnarray}}
\def\nn{\nonumber}

\begin{document}
\baselineskip=0.7cm
\setlength{\mathspace}{2.5mm}

%%%%%%%%%%%%%%%%%%%%%%%%%%%%%%%%%%%%%%%%%%%%%%%%%%%%%%%%%%

                                %titlepage

%%%%%%%%%%%%%%%%%%%%%%%%%%%%%%%%%%%%%%%%%%%%%%%%%%%%%%%%%%
\begin{titlepage}

    \begin{normalsize}
     \begin{flushright}
                 UR-1464, 
                 US-FT-20/96 \\
                 hep-th/9605073\\
     \end{flushright}
    \end{normalsize}
    \begin{LARGE}
       \vspace{1cm}
       \begin{center}
         {On M-Theory and the Symmetries of Type II }\\ 
         {String Effective Actions} \\
       \end{center}
    \end{LARGE}

  \vspace{5mm}

\begin{center}
           Ashok D{\sc as}

           \vspace{2mm}

             {\it Department of Physics and Astronomy}\\
             {\it University of Rochester,
              Rochester, N.Y. 14627, USA}\\

           \vspace{.5cm}

                \ \  and \ \

           \vspace{.5cm}

           Shibaji R{\sc oy}
           \footnote{E-mail address:
              roy@gaes.usc.es}

                 \vspace{2mm}

        {\it Departamento de F\'\i sica de Part\'\i culas} \\
        {\it Universidade de Santiago,
         E-15706 Santiago de Compostela, Spain}\\
      \vspace{1cm}

    \begin{large} ABSTRACT \end{large}
        \par
\end{center}
 \begin{normalsize}
\ \ \ \
We study the ``ordinary'' Scherk-Schwarz dimensional reduction of the bosonic
sector of the low energy effective action of a hypothetical M-theory on
$S^1 \times S^1 \cong T^2$. We thus obtain the low energy effective actions
of type IIA string theory in both ten and nine space-time dimensions. 
We point out how to obtain the O(1, 1) invariance of the NS-NS sector of
the string effective action correctly in nine dimensions. We dimensionally
reduce the type IIB string effective action  on $S^1$ and
show that the resulting nine dimensional theory can be mapped, 
purely from the bosonic
consideration, exactly to the type IIA theory by an O(1, 1) or Buscher's 
T-duality transformations. We then give a dynamical argument, in analogy with
that for the type IIB theory in ten dimensions, to show how an S-duality in 
the type IIA theory can be
understood from the underlying nine dimensional theory by compactifying 
M-theory on a T-dual torus $\tilde {T}^2$.
\end{normalsize}

\end{titlepage}
\vfil\eject
\begin{large}
\noindent{\bf I. Introduction:}
\end{large}
\vspace{.5cm}

Recently there has been a remarkable development in our understanding of the
non-perturbative behavior of various string theories. One of the most 
important new ingredients in the dynamics of various string theories is 
their close relationship [1--3] with the eleven dimensional 
supergravity theory
of Cremmer, Julia and Scherk (CJS) [4, 5]. This latter theory 
is believed to be 
the low energy description of an eleven dimensional hypothetical quantum
theory which is non-commitally called as M-theory [6] in the literature. It is
rather surprising that all the five known string theories including type I,
two type II and two heterotic string theories can be obtained by toroidal and
orbifold compactifications of this CJS theory [7--9]. Most intriguingly, 
the type II
string theories obtained by toroidal compactification of CJS theory 
automatically contain the non-perturbative information through the 
Ramond--Ramond
(R-R) fields. The sources for these R-R fields are believed to be 
extended objects known as Dirichlet(D)-branes [10--13] which should be 
included in the 
string theory spectrum as dictated by the underlying U-duality [14] 
symmetry of
the string theory.

In this paper, we study the ``ordinary'' dimensional reduction [15, 16] 
of the bosonic
sector of N=1, D=11 CJS supergravity theory. In this procedure one splits
the original eleven dimensional coordinates ($\check {x}^{\check {\mu}}$) 
into $D$
space-time coordinates ($x^{\mu}$) and $d$ internal coordinates ($x^m$)
\footnote[1]{Our notations and conventions are described in sec. II.} and
demands that the fields and the symmetry transformation laws of the original
theory be independent of the internal coordinates. The internal space is 
usually taken to be compact and in this case we consider it to be $S^1 \times
S^1 \cong T^2$. When we reduce the 11-dimensional supergravity theory first
on $S^1$ with
radius $R_1$, we obtain the type IIA theory in space-time dimension ten for 
small $R_1$. Since the radius of the circle is directly proportional to the 
string 
coupling constant as measured in terms of the metric of the string theory, 
the CJS 
theory represents the non-perturbative limit of the type IIA string theory. 
Type IIA
theory obtained this way, therefore, inherits non-perturbative information
from the supergravity theory which shows up in the natural appearance of the
R-R fields. Unlike the Neveu-Schwarz--Neveu-Schwarz (NS-NS) gauge fields,
the R-R fields do not couple to the ten dimensional dilaton and therefore 
remain inert under the four dimensional S-duality transformation, a natural
consequence of the fact that the R-R charges are carried by the solitonic
modes and not by the fundamental string modes [14]. We then further reduce the 
theory on a second $S^1$ with radius $R_2$ and obtain the type IIA string 
effective action in nine dimensions. It is well-known that the NS-NS sector
of a dimensionally reduced string effective action possesses a non-compact 
global O(d, d) [17] symmetry under which the moduli fields transform in a 
complicated non-linear way whereas the vector gauge fields transform linearly
as the vector representation of O(d, d) with all other fields remaining
invariant. So, the type IIA theory in nine dimensions should have a global
O(1, 1) invariance in the NS-NS sector. We here show that this is true only
if the antisymmetric Kalb-Ramond field also transforms appropriately under
the O(1, 1) transformation alongwith the moduli and the vector gauge fields.
In fact, we show that a particular combination of the two-form antisymmetric
gauge fields and the vector gauge fields remains invariant in order to recover
the O(1, 1) invariance of the full NS-NS sector of the nine dimensional 
action. This result is true in general for O(d, d) invariance of the 
dimensionally reduced string effective action. We then ask whether the whole
type IIA string effective action including the R-R sector is also invariant
under O(1, 1) transformation. We find the answer in the negative. However, 
we point out that the whole action possesses a global O(2) invariance [7] 
which
is kind of trivial since it follows directly from the Lorentz invariance
of the CJS theory in eleven dimensions.

It is known that the type IIB supergravity equations of motion [18, 19] 
can not be
obtained from a ten dimensional covariant action. However, if one sets the
self-dual five-form field strength to zero, then the equations of motion 
can be obtained from an action. We take the bosonic sector of this type IIB
string effective action which is also known to possess an exact global 
SL(2, R) invariance [20, 21]. Under this, a 
complex scalar field formed out of an R-R
scalar and the dilaton undergoes a fractional linear transformation whereas
the two two-form potentials, one from the NS-NS sector and the other from the
R-R sector, transform linearly with the other fields remaining invariant. 
We reduce this action on $S^1$ and find that the NS-NS sector of the resulting
nine-dimensional theory is O(1, 1) invariant as expected if the same 
combination of the two form gauge field and the vector gauge fields 
as found for the type IIA case remains
invariant. Now, we again ask the question what happens if we make the
O(1, 1) transformation on the full bosonic sector of the type IIB theory 
including the R-R sector.
We find that under this transformation type IIB string effective action
gets mapped precisely to the previously obtained type IIA string effective 
action in nine dimensions with some field redefinitions. In obtaining this
the fields in the type IIA would have to satisfy certain relations. 
We note that, this O(1, 1) transformation reduces precisely
to the Buscher's duality rules [22] for the various components of the metric,
the antisymmetric tensor field and the dilaton in ten dimensions.
We thus
observe, purely from bosonic considerations, that type IIA and type IIB
string theories are T-dual to each other\footnote[1]
{This was first observed in ref.[23, 10] 
from a different point of view.}.
As in type IIA theory, we point out that the nine 
dimensional type IIB theory also has a global O(2) invariance although it 
is not obtained from an eleven dimensional theory. At this point, 
it is natural to ask
whether the global SL(2, R) invariance of the ten dimensional type IIB theory
remains a symmetry of the nine dimensional theory. Naively, one would expect
this to be true since the scalar fields of the original theory remain intact
in the process of dimensional reduction. We explicitly show that this naive
expectation fails and the nine dimensional theory does not have a manifest
SL(2, R) invariance of the action. The SL(2, R) invariance seems special only
to D=10 and D=4. Finally, we note that the
SL(2, R) S-duality invariance of the type IIB theory in ten dimensions has its
origin in eleven dimensions as it could be understood by compactifying CJS 
theory on $S^1 \times S^1 \cong T^2$. We  present here an argument, in complete 
analogy with the type IIB theory [21, 24], that the type IIA theory in ten 
dimensions also possesses 
an SL(2, R) S-duality invariance which can be understood if we compactify the
CJS theory on a T-dual torus $\tilde {T}^2$. We donot yet know how this 
symmetry can be realized at
the level of the type IIA action in ten dimension.

The paper is organized as follows. In section II, we study the ``ordinary''
Scherk-Schwarz dimensional reduction of the bosonic sector of CJS 
supergravity theory on $S^1\times S^1 \cong T^2$ and obtain the low energy
effective action of the type IIA string theory in ten and nine space-time 
dimensions. We point out how to correctly recover the noncompact 
global O(1, 1) 
invariance of the NS-NS sector of the nine dimensional string effective 
action. The complete action including the R-R sector has been shown to
be invariant under a global O(2) transformation. The reduction on $S^1$
of the bosonic sector of the type IIB string effective action when the
five-form field strength is set to zero is presented in section III.
By applying an O(1, 1) transformation we show that type IIB action gets
mapped to the type IIA action when the fields in type IIA theory satisfy 
certain conditions. The nine dimensional type IIB theory is shown
not to have a manifest SL(2, R) invariance but is invariant under a global 
O(2) transformation. We also show how to understand an SL(2, R) S-duality
invariance in type IIA theory in ten dimensions by considering the
compactification of CJS theory on T-dual torus $\tilde {T}^2$. Finally, we 
present our conclusions in section IV.

\vspace{1cm}

\begin{large}
\noindent{\bf II. Dimensional Reduction of CJS Theory on $S^1\times S^1 \cong
T^2$:}
\end{large}

\vspace{.5cm}

In the first part of this section, we perform the ``ordinary'' Scherk-Schwarz
dimensional reduction [15, 16] of the bosonic 
sector of the N=1, D+d=11 supergravity
theory of Cremmer, Julia and Scherk on $S^1$ and fix our notations and
conventions. The original D+d coordinates will be split into D=10 space-time
coordinates and d=1 internal coordinate. We denote the eleven dimensional 
fields and coordinates with an `inverted hat', the ten dimensional
objects with a `hat' and objects in nine dimensions will be denoted
without `hat'. The Greek letters ($\lambda, \mu, \ldots$) in the later part of
the alphabet will denote
the curved space-time indices whereas  ($\alpha, \beta,
\ldots$) in the beginning of the alphabet will correspond to the flat tangent 
space indices. Similarly, the
latin letters ($m, n, \ldots$) represent the internal indices and
($a, b, \ldots$) will denote the corresponding tangent space
indices. The bosonic part of the CJS supergravity action has the form [5],
\begineq
S^{(11)} = \int\,d^{11}\ch x\,\ch e \left[\ch R -\frac{1}{12}
\ch F_{\ch \mu\ch \nu
\ch \rho \ch \sigma}\ch F^{\ch \mu\ch \nu \ch \rho \ch \sigma} +\frac{8}
{(12)^4}\frac{1}{\ch e}\epsilon^{\ch \mu_1\ldots \ch \mu_{11}} 
\ch F_{\ch \mu_1
\ldots \ch \mu_4}\ch F_{\ch \mu_5 \ldots \ch \mu_8} \ch C_{\ch \mu_9\ldots
\ch \mu_{11}}\right]
\endeq 
where $\ch e $ = det$(\ch e^{\ch \alpha}_{\ch \mu})$, 
$\ch e^{\ch \alpha}_{\ch \mu}$ being the elfbein, $\ch C_{\ch \mu \ch \nu
\ch \lambda}$ is an antisymmetric three-form gauge field and 
$\ch F^{\ch \mu\ch \nu \ch \rho \ch \sigma}$ is the corresponding field 
strength. $\ch R$ is the eleven dimensional scalar curvature. Our convention
for the signature of the tangent space Lorentz metric is 
$(-, +, +, \ldots)$. The scalar curvature is defined as
\begineq
\ch R = \ch e^{\ch \mu \ch \alpha}\ch e^{\ch \nu \ch \beta} \ch R_{\ch \mu
\ch \nu \ch \alpha \ch \beta}
\endeq
where
\begineq
\ch R_{\ch \mu \ch \nu \ch \alpha \ch \beta} = \partial_{\ch \mu} 
\ch \omega_{\ch \nu \ch \alpha \ch \beta} + \ch \omega_{\ch \mu \ch \alpha}
^{\,\,\,\,\,\,\ch \gamma} \ch \omega_{\ch \nu \ch \gamma \ch \beta} - (\ch \mu
\leftrightarrow \ch \nu)
\endeq
Our convention for the spin connection is
\begineq
\ch \omega_{\ch \alpha \ch \beta \ch \gamma} = - \ch \Omega_{\ch \alpha \ch
\beta, \ch \gamma} + \ch \Omega_{\ch \beta \ch \gamma, \ch \alpha} - 
\ch \Omega_{\ch \gamma \ch \alpha, \ch \beta} 
\endeq
with
\begineq
\ch \Omega_{\ch \alpha \ch \beta, \ch \gamma} = \frac{1}{2}\left(
\ch e_{\ch \alpha}^{\,\,\,\ch \mu} \partial_{\ch \beta} 
\ch e_{\ch \gamma \ch \mu}
- \ch e_{\ch \beta}^{\,\,\,\ch \mu} \partial_{\ch \alpha} 
\ch e_{\ch \gamma \ch \mu}\right)
\endeq
We note that $\ch \Omega_{\ch \alpha \ch \beta, \ch \gamma}$ is antisymmetric
in its first two indices whereas $\ch \omega_{\ch \alpha \ch 
\beta \ch \gamma}$ is antisymmetric in its last two indices. The field 
content of the theory (1) is just an elfbein $\ch e_{\ch \mu}^{\,\,\,
\ch \alpha}$
and a totally antisymmetric three-form potential $\ch C_{\ch \mu \ch \nu 
\ch \rho}$. The scalar curvature term in (1) will be simplified by using the
following identity which is valid in any space-time dimensions:
\eqabegin
& &\int\,d^D x e \Lambda(x) R\nn\\
\qquad &=& \int\,d^D x e \Lambda(x)\left[\omega_{\alpha\beta\gamma}\omega^
{\alpha\beta\gamma} + \omega^{\alpha}_{\,\,\,\alpha\gamma}\omega^{\beta
\,\,\,\,\gamma}_{\,\,\,\beta} + 2 e^{\mu\alpha}(\partial_\mu \log \Lambda)
\omega^{\beta}_{\,\,\,\beta\alpha}\right]
\eqaend
Where $\Lambda(x)$ is an arbitrary function of $x$. 

In the ``ordinary'' dimensional reduction the field variables will be taken 
as independent of the internal coordinates. Using SO(1, 10) Lorentz invariance
of the eleven dimensional theory the elfbein is usually taken in the 
triangular form [5] as given below:
\begineq
\ch e_{\ch \mu}^{\,\,\,\ch \alpha} = \left(\begin{array}{cc}
e_\mu^{\,\,\alpha}  & A_\mu^{\,n} e_n^{\,\,a} \\
                0   &     e_m^{\,\,a}\end{array}\right)
\qquad {\rm and}\quad 
\ch e_{\ch \alpha}^{\,\,\,\ch \mu} = \left(\begin{array}{cc}
e_\alpha^{\,\,\mu}  & - e_\alpha^{\,\,\nu}A_\nu^{\,m} \\
                0   &     e_a^{\,\,m}\end{array}\right)
\endeq
With this convention the metric and its inverse then take the following form:
\begineq
\ch g_{\ch \mu \ch \nu} = \left(\begin{array}{cc}
g_{\mu\nu} + A_\mu^{\,m} A_{\nu m}  & A_{\mu n}  \\
       A_{\nu m}            &     g_{mn}\end{array}\right)
\qquad {\rm and}\quad
\ch g^{\ch \mu \ch \nu} = \left(\begin{array}{cc}
g^{\mu\nu}  & - A^{\,\nu m} \\
  -A^{\, \mu n}                 &     A_\mu^{\, n}
A^{\,\mu p} + g^{np}\end{array}\right)
\endeq
We have not used any accent to denote the fields on the right hand side 
because we will
take these structures of the vielbein and the metric for both ten and nine
dimensions. Here $e_\mu^{\,\,\alpha}$ and $e_m^{\,\,a}$ are respectively 
the D(space-time) and d(internal) vielbeins. $A_{\mu n}$ are the d vector
gauge fields which result from the dimensional reduction. In this convention,
the non-vanishing components of the spin-connections are:
\eqabegin
\omega_{\alpha\beta\gamma} & &\nn\\
\omega_{\alpha\beta a} &=& \frac{1}{2} e_\alpha^{\,\,\mu} e_{\beta}^{\,\,\nu}
e_{a n} F_{\mu\nu}^{\,n} = - \omega_{a \alpha\beta}\nn\\
\omega_{\alpha a b} &=& \frac{1}{2}\left[e_a^{\,\,m} e_{\alpha}^{\,\,\mu} 
\partial_\mu e_{b m} - e_b^{\,\,m} e_{\alpha}^{\,\,\mu} 
\partial_\mu e_{a m}\right]\\
\omega_{a \alpha b} &=& -\frac{1}{2} e_a^{\,\,m} e_b^{\,\,n} 
e_\alpha^{\,\,\mu}\partial_\mu g_{mn}\nn
\eqaend
where $F_{\mu\nu}^{\,m} = \partial_\mu A_\nu^{\,m} - 
\partial_\nu A_\mu^{\,m}$. Using (6) with $\Lambda(x) = 1$, (7) and (9)
it is a straightforward exercise to verify the following relation for the 
dimensional reduction from 11 $\rightarrow$ D dimensions (note that this
relation is valid in general for the dimensional reduction from any
D$+$d $\rightarrow $ D dimensions).
\eqabegin
\int\,d^{11} \ch x \ch e \ch R         
&\rightarrow &\int d^D x \sqrt {-g} \Delta \left[R - \frac{1}{4}
g_{mn} F_{\mu\nu}^{\,m} F^{\,\mu\nu n}\right.\nn\\ 
& &\qquad\left.+ \frac{1}{4} g^{\mu\nu}
\partial_{\mu} g_{mn} \partial_\nu g^{mn} + g^{\mu\nu}\partial_\mu \log\Delta
\partial_\nu\log\Delta\right]
\eqaend
where $g$ = det $g_{\mu\nu}$, $\Delta^2$ = det $g_{mn}$, $g_{\mu\nu}$ and
$R$ being the D dimensional metric and the scalar curvature respectively. 
In (10) we 
have also set the integral over the internal coordinates on a compact
manifold to unity. Since
we want to reduce the theory (1) just by one dimension we choose $m=n=10$
and
\eqabegin
\Delta^2 &=& {\rm det} g_{mn} = g_{10,10}\nn\\
\h A^{10}_{\h \mu} &\equiv & \h A^{(1)}_{\h \mu}
\eqaend
Using this $g_{10,10}$ and $\h A^{10}_{\h \mu}$ in eq.(10) and 
performing the dimensional reductions of
the other terms in (1) we obtain,
\eqabegin
S^{(10)} &=& \int\,d^{10}\h x \h e \left[\Delta \h R - \frac{1}{4} \Delta^3
\h F_{\h \mu \h \nu}^{(1)} \h F^{(1)\,\h \mu \h \nu} - \frac{1}{3} \Delta^{-1}
\h H_{\h \mu \h \nu \h \rho}^{(1)} \h H^{(1)\,\h \mu \h \nu \h \rho}
-\frac{1}{12}\Delta \h F_{\h \mu \h \nu \h \rho \h \sigma}
\h F^{\h \mu \h \nu \h \rho \h \sigma}\right.\nn\\ 
& &\qquad\left.+\frac{8}{(12)^4}\frac{1}{\h e}\epsilon^{\h \mu_1\ldots
\h \mu_{10}}\left(3 \h F_{\h \mu_1 \ldots
\h \mu_4} \h F_{\h \mu_5\ldots \h \mu_8} \h B^{(1)}_{\h \mu_9 
\h \mu_{10}} - 8 \h F_{\h \mu_1 \ldots\h \mu_4} \h H_{\h \mu_5\h \mu_6
\h \mu_7}^{(1)} \h C_{\h \mu_8 \h \mu_9\h \mu_{10}}\right)\right]
\eqaend
where we have defined,
\eqabegin
\h B_{\h \mu \h \nu}^{(1)} &=& \ch C_{\ch \mu \ch \nu 10} \,\,\,=\,\,\,\h C_{
\h \mu \h \nu 10}\nn\\
\h C_{\h \mu \h \nu \h \rho} &=& \ch C_{\h \mu \h \nu \h \rho}-\left(
\h A_{\h \mu}^{(1)} \h B_{\h \nu \h \rho}^{(1)} + {\rm cyc.\,\, in}\,\,\, 
\h \mu \h \nu \h \rho\right)\nn\\
\h H_{\h \mu \h \nu \h \rho}^{(1)} &=& \partial_{\h \mu} \h B_{\h \nu \h \rho}
^{(1)} + {\rm cyc.\,\, in}\,\,\, \h \mu \h \nu \h \rho\\
\h F_{\h \mu \h \nu \h \rho \h \sigma} &=& 
\partial_{\h \mu} \h C_{\h \nu \h \rho \h \sigma} - 
\partial_{\h \nu} \h C_{\h \mu \h \rho \h \sigma} +
\partial_{\h \rho} \h C_{\h \sigma \h \mu \h \nu} -
\partial_{\h \sigma} \h C_{\h \rho \h \mu \h \nu}\nn\\
& & +\left(\h F_{\h \mu \h \nu}^{(1)} \h B_{\h \rho \h \sigma}^{(1)}
    + \h F_{\h \nu \h \rho}^{(1)} \h B_{\h \mu \h \sigma}^{(1)} + {\rm cyc.
\,\, in}\,\,\,\h \nu \h \rho \h \sigma\right)\nn
\eqaend
We notice in (12) that the scalar curvature term and the $(\h H^{(1)})^2$ 
term
have different powers of $\Delta$. In order to reproduce the NS-NS sector
of the type IIA string effective action correctly, they should have the same
power of $\Delta$ which could then be identified with the usual
dilaton coupling. We notice that this could be done with the rescaling of
the ten-dimensional metric by
\begineq
\h g_{\h \mu \h \nu} \rightarrow \Delta^{-1} \h g_{\h \mu \h \nu}
\endeq
With this rescaling we find that both the terms are multiplied by 
$\Delta^{-3}$. So, in order to produce the correct dilaton coupling we
set
\eqabegin
\Delta^{-3} &=& e^{-2\h \phi}\nn\\
{\rm or,}\,\,\,\, \Delta &=& e^{\frac{2}{3}\h \phi}
\eqaend
We also note that for a general rescaling of the metric of the form 
$g_{\mu\nu} \rightarrow e^{\alpha \phi(x)} g_{\mu\nu}$, in D dimensions, the
scalar curvature changes as 
\begineq
R\rightarrow e^{-\alpha \phi}\left[R - \alpha\left(D-1\right)g^{\mu\nu}
\nabla_\mu\nabla_\nu\phi - \frac{\alpha^2}{4}\left(D-1\right)\left(D-2\right)
g^{\mu\nu}\nabla_\mu\phi\nabla_\nu\phi\right]
\endeq
So, under this rescaling $\sqrt{-g} R$ changes as
\eqabegin
& & \sqrt{-g} R\nn\\
&\rightarrow & \sqrt{-g} e^{\frac{1}{2} (D-2) \alpha \phi}
\left[R - \alpha\left(D-1\right)g^{\mu\nu}
\nabla_\mu\nabla_\nu\phi - \frac{\alpha^2}{4}\left(D-1\right)\left(D-2\right)
g^{\mu\nu}\nabla_\mu\phi\nabla_\nu\phi\right]\nn
\eqaend
Using (16), (15) and (14) in (12), we obtain the type IIA string effective
action in D=10 in the form,
\eqabegin
S^{(10)}_{{\rm IIA}} &=& \int\, d^{10} \h x \sqrt{-\h g} \left[ 
e^{-2\h \phi}\left(\h R + 4 \partial_{\h \mu}\h \phi \partial^{\h \mu}\h \phi
-\frac{1}{3} \h H_{\h \mu \h \nu \h \rho}^{(1)}
\h H^{(1)\,\h \mu \h \nu \h \rho}\right) - \frac{1}{4} \h F_{\h \mu \h \nu}
^{(1)}\h F^{(1)\,\h \mu \h \nu}\right.\nn\\
& &\qquad\qquad -\frac{1}{12} \h F_{\h \mu \h \nu \h \rho \h \sigma}
\h F^{\h \mu \h \nu \h \rho \h \sigma} + \frac{8}{(12)^4}\frac{\epsilon^
{\h \mu_1 \ldots \h \mu_{10}}}{\sqrt {-\h g}}\left(
3\h F_{\h \mu_1\ldots \h \mu_4}
\h F_{\h \mu_5\ldots \h \mu_8} \h B_{\h \mu_9 \h \mu_{10}}^{(1)}\right.\nn\\
& &\qquad\qquad \left.\left.- 8 \h F_{\h \mu_1 \ldots\h \mu_4} 
\h H_{\h \mu_5\h \mu_6
\h \mu_7}^{(1)} \h C_{\h \mu_8 \h \mu_9\h \mu_{10}}\right)\right]
\eqaend
We here note that the first three terms in (17) with the usual dilaton
coupling represent the NS-NS sector and is common to all string theories.
The fourth and the fifth terms belong to the R-R sector whereas the last
term has a mixing between the NS-NS and R-R sectors.
The last three terms, however, do not couple to the dilaton  
and encode the non-perturbative information inherited from
CJS theory in eleven dimensions. Also, we note from (15) that as the 
expectation value of $\Delta$ represents the radius of compactification, $R_1$,
of the circle $S^1$, 
it is related to the string coupling constant as 
$R_1 \sim e^{\frac{2}{3}\h \phi} \sim \lambda_{{\rm A}}^{\frac{2}{3}}$. 
Because of the Weyl scaling (14), the radius as measured by the string metric
would be $R_1^s \sim e^{\frac{1}{3}\h \phi} e^{\frac{2}{3}\h \phi} \sim
\lambda_{{\rm A}}$. So, as $R_1^s \rightarrow \infty,\,\,\,\lambda_{{\rm A}}
\rightarrow \infty$ and therefore, CJS theory represents the non-perturbative 
limit of type IIA theory [1].

In the second part of this section, we consider the dimensional reduction of
the
type IIA theory, eq.(17), on a second $S^1$. In this case we follow exactly
the same procedure as before and take the zehnbein and the ten dimensional
metric in the same form as given in (7) and (8). We take the ninth component 
of the
metric  as $g_{99} = \chi$ and the vector gauge field which originates 
in the dimensional reduction of the ten dimensional metric $\h g_{\h \mu
\h \nu}$ is denoted as $A_{\mu}^{(2)}$. In order to reduce the scalar 
curvature and the dilaton term in (17) we note the following relation for
the dimensional reduction from 10 $\rightarrow$ D dimensions (this relation
is also valid for the reduction from any D+d $\rightarrow$ D),
\eqabegin
& &\int\,d^{10} \h x \sqrt {- \h g} e^{-2\h \phi}\left(\h R + 
4 \partial_{\h \mu}\h \phi\partial^{\h \mu}\h \phi\right)\nn\\
& & \rightarrow \int\, d^D x \sqrt {- g} e^{-2\phi_D}\left(R + 
4 \partial_\mu\phi_D\partial^\mu\phi_D - \frac{1}{4}g_{mn}F_{\mu\nu}^m
F^{\mu\nu,n}+ \frac{1}{4} \partial_\mu g_{mn}\partial^\mu g^{mn}\right)
\eqaend
where D is any dimension lower than ten. Note that we have here used the
identity (6) with $\Lambda = e^{-2\h \phi}$. The dilaton fields $\h \phi$
and $\phi_D$ are related as,
\begineq
\h \phi = \phi_D + \frac{1}{4} \log\left({\rm det} g_{mn}\right)
\endeq
For $m=n=9$ and $A_\mu^9 \equiv A_\mu^{(2)}$, we obtain from (18) the reduced
form of the scalar curvature and the dilaton term as (We will denote the nine
dimensional dilaton as $\phi$.),
\begineq
\int\, d^9 x \sqrt{-g} e^{-2\phi}\left(R+4\partial_\mu\phi\partial^\mu\phi
-\frac{1}{4}\partial_\mu\log\chi\partial^\mu\log\chi -
\frac{1}{4}\chi F_{\mu\nu}^{(2)} F^{(2)\,\mu\nu}\right)
\endeq
We then write below the reduced form of the other terms in (17) separately,

\noindent{3rd term:}
\eqabegin
& & \int\,d^{10} \h x \sqrt {- \h g} e^{-2 \h \phi} \left(-\frac{1}{3}\right)
\h H_{\h \mu\h \nu\h \rho}^{(1)}\h H^{(1)\,\h \mu\h \nu\h \rho}\nn\\
&\rightarrow &\int d^9 x \sqrt{-g} e^{-2\phi}\left(-\frac{1}{3}
H_{\mu \nu \rho}^{(1)} H^{(1)\, \mu \nu \rho} - \chi^{-1} 
F_{\mu\nu}^{(3)} F^{(3)\,\mu\nu}\right)
\eqaend
\noindent{4th term:}
\eqabegin
& &\int\,d^{10} \h x \sqrt{-\h g} \left(-\frac{1}{4}\right)
\h F_{\h \mu\h \nu}^{(1)} \h F^{(1)\,\h \mu\h \nu}\nn\\
&\rightarrow &\int d^9 x \sqrt{-g}\left[-\frac{1}{4}\chi^{\frac{1}{2}}\left(
F_{\mu\nu}^{(1)} + a F_{\mu\nu}^{(2)}\right)\left(F^{(1)\,\mu\nu} + a 
F^{(2)\,\mu\nu}\right) - \frac{1}{2} \chi^{-\frac{1}{2}}\partial_\mu a
\partial^\mu a\right]
\eqaend
\noindent{5th term:}
\eqabegin
& &\int\,d^{10} \h x \sqrt{-\h g} \left(-\frac{1}{12}\right)
\h F_{\h \mu\h \nu \h \rho \h \sigma}\h F^{\h \mu\h \nu \h \rho \h \sigma}\nn\\
&\rightarrow &\int d^9 x \sqrt{-g}\left[-\frac{1}{12}\chi^{\frac{1}{2}}
F_{\mu \nu \rho \sigma}F^{\mu \nu \rho \sigma} - \frac{1}{3} 
\chi^{-\frac{1}{2}}\left(H_{\mu \nu \rho}^{(2)} - a H_{\mu \nu \rho}^{(1)}
\right)\left(H^{(2)\,\mu \nu \rho} - a H^{(1)\,\mu \nu \rho}\right)\right]
\nn\\
\eqaend
\noindent{last term:}
\eqabegin
& & \int\, d^{10} \h x \frac{8}{(12)^4}\epsilon^{\h \mu_1\ldots
\h \mu_{10}}\left(3 \h F_{\h \mu_1 \ldots
\h \mu_4} \h F_{\h \mu_5\ldots \h \mu_8} \h B^{(1)}_{\h \mu_9
\h \mu_{10}} - 8 \h F_{\h \mu_1 \ldots\h \mu_4} \h H_{\h \mu_5\h \mu_6
\h \mu_7}^{(1)} \h C_{\h \mu_8 \h \mu_9\h \mu_{10}}\right)
\nn\\
&\rightarrow & \int\,d^9 x \frac{\epsilon^{\mu_1\ldots \mu_9}}
{3 (12)^2}\left[F_{\mu_1
\ldots \mu_4} F_{\mu_5\ldots \mu_8} A_{\mu_9}^{(3)} - 4 F_{\mu_1\ldots \mu_4}
\epsilon^{ij} H_{\mu_5\mu_6\mu_7}^{(i)}\bar{B}_{\mu_8\mu_9}^{(j)}\right.\nn\\
& & \qquad + 4 F_{\mu_1\ldots \mu_4} H_{\mu_5\mu_6\mu_7}^{(i)} A_{\mu_8}^{(i)}
A_{\mu_9}^{(3)} + 2 \epsilon^{ij} H_{\mu_1\mu_2\mu_3}^{(i)} 
H_{\mu_4\mu_5\mu_6}^{(j)} \bar{C}_{\mu_7\mu_8\mu_9}\nn\\
& &\qquad + 4 F_{\mu_1\ldots \mu_4}F_{\mu_5\mu_6}^{(3)}
\bar{C}_{\mu_7\mu_8\mu_9} + 6 \epsilon^{ij} H_{\mu_1\mu_2\mu_3}^{(i)}
H_{\mu_4\mu_5\mu_6}^{(j)}A_{\mu_7}^{(k)}\bar{B}_{\mu_8\mu_9}^{(k)}\nn\\
& &\qquad \left.+ 12 F_{\mu_1\ldots \mu_4}F_{\mu_5\mu_6}^{(3)}A_{\mu_7}^{(i)}
\bar{B}_{\mu_8\mu_9}^{(i)}\right]
\eqaend
We have expressed the reduced form of this term in terms of the indexed fields
for later convenience, where $i$, $j$, $k$ = 1, 2 and $\epsilon^{12} =
-\epsilon^{21} = 1$.
Our definitions and the dimensionally reduced form of the gauge fields as well
as their field strengths are listed below:
\eqabegin
\h g_{\h \mu\h \nu}&\longrightarrow & \left\{ \begin{array}{l} \h g_{99} 
 = g_{99} = \chi
\\
\h g_{\mu 9} = g_{\mu 9} = \chi A_\mu^{(2)} \\
\h g_{\mu\nu}= g_{\mu\nu} + \chi A_\mu^{(2)} A_\nu^{(2)} \end{array}\right.\\
\h \phi &=& \phi + \frac{1}{4} \log \chi\\
\h A_{\h \mu}^{(1)} & \longrightarrow & \left\{ \begin{array}{l} 
A_9 = \h A_9 = a \\
A_{\mu}^{(1)}
= \h A_{\mu}^{(1)} - a A_{\mu}^{(2)} 
\end{array}\right.\\
\h B_{\h \mu \h \nu}^{(1)} &\longrightarrow & \left\{ \begin{array}{l}
B_{\mu 9}^{(1)} = \h B_{\mu 9}^{(1)} = A_\mu^{(3)} \\
B_{\mu\nu}^{(1)}
= \h B_{\mu\nu}^{(1)} + A_{\mu}^{(2)} A_{\nu}^{(3)} - 
A_{\nu}^{(2)} A_{\mu}^{(3)} 
\end{array}\right.\\
\h C_{\h \mu \h \nu \h \lambda} &\longrightarrow &\left\{ \begin{array}{l}
C_{\mu \nu 9} = \h C_{\mu\nu 9} = B_{\mu\nu}^{(2)} - a B_{\mu\nu}^{(1)}
-\left(A_{\mu}^{(1)} A_{\nu}^{(3)} -
A_{\nu}^{(1)} A_{\mu}^{(3)}\right) \\
C_{\mu\nu\lambda}
= \h C_{\mu\nu\lambda} - \left(A_{\mu}^{(2)} \h C_{\nu\lambda 9} + {\rm cyc.
\,\, in}\,\,\, \mu\nu\lambda\right)
\end{array}\right.
\eqaend
The field strength associated with various gauge fields are given below:
\eqabegin
F_{\mu\nu}^{(2)} &=& \partial_\mu A_\nu^{(2)} - \partial_\nu A_\mu^{(2)}\\
F_{\mu\nu}^{(3)} &=& \partial_\mu A_\nu^{(3)} - \partial_\nu A_\mu^{(3)}\\
F_{\mu 9} &=& \partial_\mu a\\
F_{\mu\nu} &=& F_{\mu\nu}^{(1)} + a F_{\mu\nu}^{(2)}\\
H_{\mu\nu 9}^{(1)} &=& \h H_{\mu\nu 9}^{(1)}\,\,\,=\,\,\,F_{\mu\nu}^{(3)}\\
H_{\mu\nu \lambda}^{(1)} &=& \partial_\mu B_{\nu\lambda}^{(1)} - 
F_{\mu\nu}^{(2)} A_\lambda^{(3)} + {\rm cyc.\,\, in}\,\,\,\mu\nu\lambda\nn\\
&=& \partial_\mu \bar{B}_{\nu\lambda}^{(1)} -\frac{1}{2}\left(
F_{\mu\nu}^{(2)} A_\lambda^{(3)} + F_{\mu\nu}^{(3)} A_\lambda^{(2)}\right) 
+ {\rm cyc.\,\, in}\,\,\,\mu\nu\lambda
\eqaend
where we have defined, 
\begineq
\bar{B}_{\mu\nu}^{(1)} \equiv B_{\mu\nu}^{(1)} - \frac{1}{2}\left(
A_{\mu}^{(2)} A_\nu^{(3)} - A_{\nu}^{(2)} A_\mu^{(3)}\right)
\endeq
Continuing with the other field strengths,
\begineq
F_{\mu\nu\lambda 9} = \h F_{\mu\nu\lambda 9} = H_{\mu\nu\lambda}^{(2)}
- a H_{\mu\nu\lambda}^{(1)}
\endeq
where $H_{\mu\nu\lambda}^{(2)}$ is defined as
\eqabegin
H_{\mu\nu \lambda}^{(2)} &=& \partial_\mu B_{\nu\lambda}^{(2)} +
F_{\mu\nu}^{(3)} A_\lambda^{(1)} + {\rm cyc.\,\, in}\,\,\,\mu\nu\lambda\nn\\
&=& \partial_\mu \bar{B}_{\nu\lambda}^{(2)} +\frac{1}{2}\left(
F_{\mu\nu}^{(1)} A_\lambda^{(3)} + F_{\mu\nu}^{(3)} A_\lambda^{(1)}\right)
+ {\rm cyc.\,\, in}\,\,\,\mu\nu\lambda
\eqaend
where,
\begineq
\bar{B}_{\mu\nu}^{(2)} \equiv B_{\mu\nu}^{(2)} -\frac{1}{2}\left(
A_{\mu}^{(1)} A_\nu^{(3)} - A_{\nu}^{(1)} A_\mu^{(3)}\right)
\endeq
and finally,
\eqabegin
F_{\mu\nu\lambda\rho} &=& \partial_\mu \bar{C}_{\nu\lambda\rho} - 
\partial_\nu \bar{C}_{\mu\lambda\rho} + \partial_\lambda 
\bar{C}_{\rho\mu\nu} - \partial_\rho \bar{C}_{\lambda\mu\nu}\nn\\
&+&\left[F_{\mu\nu}^{(i)}\bar{B}_{\lambda\rho}^{(i)} + F_{\nu\lambda}^{(i)}
\bar{B}_{\mu\rho}^{(i)} - \frac{1}{2}\epsilon^{ij} F_{\mu\nu}^{(3)} A_\lambda
^{(i)} A_\rho^{(j)} - \frac{1}{2}\epsilon^{ij} F_{\nu\lambda}^{(3)} A_\mu
^{(i)} A_\rho^{(j)} + {\rm cyc.\,\, in}\,\,\,\,\nu\lambda\rho\right]\nn\\
\eqaend
where,
\begineq 
\bar{C}_{\mu\nu\lambda} = C_{\mu\nu\lambda} + \left[\frac{1}{2} \epsilon^{ij}
A_{\mu}^{(i)} A_{\nu}^{(j)} A_{\lambda}^{(3)} + {\rm cyc.\,\, in}\,\,\,\,
\mu\nu\lambda\right]
\endeq
Note from (25) that the radius of the circle of compactification is given by
the expectation value of the field $\chi^{\frac{1}{2}}$ as measured in string 
metric and will be used later. Also, we have introduced both $B_{\mu\nu}$
and $\bar{B}_{\mu\nu}$, because we will see that it is $\bar{B}_{\mu\nu}$ which
will remain invariant under the global O(1, 1) transformation and not 
$B_{\mu\nu}$. Under the global O(2) transformations also, $\bar{B}_{\mu\nu}$'s
transform in a nice way as opposed to $B_{\mu\nu}$'s. The particular forms of 
the field strengths are chosen for later convenience.

The complete reduced form of the type IIA string effective action in nine 
dimensions can be written using (20)--(24) as follows:
\eqabegin
S^{(9)}_{{\rm IIA}} &=&
\int\,d^{9}x\sqrt {-g} \left[ e^{-2\phi}\left(R + 4
\partial_\mu \phi\partial^\mu \phi -\frac{1}{4} \partial_\mu \log \chi
\partial^\mu \log \chi  -\frac{1}{4} \chi
F_{\mu\nu}^{(2)} F^{(2)\,\mu\nu}\right.\right.\nn\\
& &\qquad\qquad \left. 
- \frac{1}{4}\chi^{-1} F_{\mu\nu}^{(3)}
F^{(3)\,\mu\nu} -\frac{1}{12} H_
{\mu\nu\lambda}^{(1)} H^{(1)\,\mu\nu\lambda}
\right) -\frac{1}{2} \chi^{-\frac{1}{2}}\partial_\mu a
\partial^\mu a\nn\\
& &\qquad\qquad -\frac{1}{12}\chi^{-\frac{1}{2}}
\left(H_{\mu\nu\lambda}^{(2)} - a H_{\mu\nu
\lambda}^{(1)}\right)\left(H^{(2)\,\mu\nu\lambda}
-a H^{(1)\,\mu\nu\lambda}
\right)\nn\\
& &\qquad\qquad -\frac{1}{4} \chi^{\frac{1}{2}}
\left(F_{\mu\nu}^{(1)} + a F_{\mu\nu}^{(2)}
\right)\left(F^{(1)\,\mu\nu} + a F^{(2)\,\mu\nu}\right) -
\frac{1}{12}\chi^{\frac{1}{2}}F_{\mu\nu\lambda\rho}
F^{\mu\nu\lambda\rho}\nn\\ 
& &\qquad +\frac{\epsilon^{\mu_1\ldots \mu_9}}{\sqrt {-g}}\frac{1}
{2 (12)^3}\left(F_{\mu_1
\ldots \mu_4} F_{\mu_5\ldots \mu_8} A_{\mu_9}^{(3)} - 4 F_{\mu_1\ldots \mu_4}
\epsilon^{ij} H_{\mu_5\mu_6\mu_7}^{(i)}\bar{B}_{\mu_8\mu_9}^{(j)}\right.\nn\\
& & \qquad\qquad + 4 F_{\mu_1\ldots \mu_4} H_{\mu_5\mu_6\mu_7}^{(i)} 
A_{\mu_8}^{(i)}
A_{\mu_9}^{(3)} + 2 \epsilon^{ij} H_{\mu_1\mu_2\mu_3}^{(i)}
H_{\mu_4\mu_5\mu_6}^{(j)} \bar{C}_{\mu_7\mu_8\mu_9}\nn\\
& &\qquad\qquad + 4 F_{\mu_1\ldots \mu_4}F_{\mu_5\mu_6}^{(3)}
\bar{C}_{\mu_7\mu_8\mu_9} + 6 \epsilon^{ij} H_{\mu_1\mu_2\mu_3}^{(i)}
H_{\mu_4\mu_5\mu_6}^{(j)}A_{\mu_7}^{(k)}\bar{B}_{\mu_8\mu_9}^{(k)}\nn\\
& &\qquad\qquad \left.\left.+ 12 F_{\mu_1\ldots \mu_4}F_{\mu_5\mu_6}^{(3)}
A_{\mu_7}^{(i)}
\bar{B}_{\mu_8\mu_9}^{(i)}\right)\right]
\eqaend
In writing down (42) we have rescaled $A_\mu^{(3)}\rightarrow \frac{1}{2}
A_\mu^{(3)}$, $B_{\mu\nu}^{(1)}\rightarrow \frac{1}{2} B_{\mu\nu}^{(1)}$,
$B_{\mu\nu}^{(2)}\rightarrow \frac{1}{2} B_{\mu\nu}^{(2)}$ 
and $C_{\mu\nu\lambda} \rightarrow \frac{1}{2} C_{\mu\nu\lambda}$ to recast
the action in the standard form. This action 
is also obtained in ref.[7] with different definitions and conventions of the
reduced fields. But, note that we do not agree with the reduction of the
topological term eq.(47) in ref.[7]. 
The first six terms in (42) represent the NS-NS sector which 
couples to nine dimensional dilaton $\phi$, the next four terms 
which do not couple to the dilaton encode 
the non-perturbative information and represent the R-R sector whereas
the last term gives a mixing between the NS-NS and the R-R sectors. We first 
concentrate on the NS-NS sector of the action (42) and show that it is O(1, 1)
invariant. We here follow closely the notation adopted in the paper of
Maharana and Schwarz [15]. Since the nine 
dimensional dilaton and the metric remain 
invariant under O(1, 1) transformation the first two terms are invariant.
The third, fourth and fifth terms can be written as
\eqabegin
& &-\frac{1}{4} \partial_\mu \log \chi
\partial^\mu \log \chi  -\frac{1}{4} \chi
F_{\mu\nu}^{(2)} F^{(2)\,\mu\nu}
- \frac{1}{4}\chi^{-1} F_{\mu\nu}^{(3)}
F^{(3)\,\mu\nu}\nn\\
& &\qquad\qquad =\,\,\, \frac{1}{8}{\rm tr}\,\,
\partial_{\mu}M\partial^\mu M^{-1} - \frac{1}{4}
{\cal F}_{\mu\nu}^T M^{-1} {\cal F}^{\mu\nu}
\eqaend
where
\begineq
M = \left(\begin{array}{cc}
\chi^{-1}  &  0 \\
       0   &  \chi\end{array}\right)
\qquad {\rm and}\qquad
{\cal F}_{\mu\nu} = \left(\begin{array}{c}
F_{\mu\nu}^{(2)}\\ F_{\mu\nu}^{(3)}
\end{array}\right)
\endeq
Here $M$ is an O(1, 1) matrix, since $M^T \eta M = \eta$ with 
$\eta = \left(\begin{array}{cc}
 0 &  1 \\
       1   & 0\end{array}\right)$. So, (43) is invariant under a global 
O(1, 1) transformation $M\rightarrow \Omega M \Omega^T$, where $\Omega^T
\eta \Omega = \eta$, if ${\cal A}_{\mu} = \left(\begin{array}{c}
A_{\mu}^{(2)}\\ A_{\mu}^{(3)}
\end{array}\right)$ transforms as ${\cal A}_\mu \rightarrow \Omega 
{\cal A}_\mu$. The form of the O(1, 1) transformation matrix 
$\Omega$ is:
\begineq
\Omega = \left(\begin{array}{cc}
0  &  \lambda \\
\lambda^{-1}   &  0\end{array}\right)
\endeq
where $\lambda$ is a constant parameter. So, we find the O(1, 1) 
transformation 
of the moduli $\chi$ and the vector gauge fields as
\eqabegin
\tilde {\chi} &=& \lambda^{-2} \chi^{-1}\nn\\
\tilde {A}_\mu^{(2)} &=& \lambda A_\mu^{(3)}\\
\tilde {A}_\mu^{(3)} &=& \lambda^{-1} A_\mu^{(2)}\nn
\eqaend
We note that (43) is indeed invariant under the transformations in (46). 
Next we find that $(H^{(1)})^2$ term in (42) is also invariant under O(1, 1)
transformation since from eq.(35) we observe that $H^{(1)}_{\mu\nu\lambda}$
can be expressed as
\begineq
H_{\mu\nu\lambda}^{(1)} = \partial_\mu \bar{B}_{\nu\lambda}^{(1)} -
\frac{1}{2} {\cal A}_\mu^T \eta {\cal F}_{\nu\lambda} + {\rm cyc.\,\,in}\,\,
\,\,\mu\nu\lambda
\endeq
and this is invariant if we require that $\bar{B}_{\mu\nu}^{(1)} = 
B_{\mu\nu}^{(1)} - \frac{1}{2}\left(A_\mu^{(2)}A_\nu^{(3)}-
A_\nu^{(2)}A_\mu^{(3)}\right)$ does not transform. The second term in (47) is
invariant because $\Omega^T\eta\Omega =\eta$. This shows that it is 
$\bar{B}_{\mu\nu}^{(1)}$ and not $B_{\mu\nu}^{(1)}$ which should remain 
invariant under O(1, 1) (as noted incorrectly in ref.[17])  
for the O(1, 1)
invariance of the NS-NS sector. In fact, $B_{\mu\nu}^{(1)}$ in the present 
case transforms under (46) as,
\begineq
\tilde{B}_{\mu\nu}^{(1)} = B_{\mu\nu}^{(1)} -\left(A_\mu^{(2)}A_\nu^{(3)}-
A_\nu^{(2)}A_\mu^{(3)}\right)
\endeq
When both $A_\mu^{(2)}$ and $A_\mu^{(3)}$ are chosen to be zero, it is only 
then $B_{\mu\nu}^{(1)}$ would remain invariant. We will see in section III 
that the transformation (46) corresponds to Buscher's duality transformation
for a special value of $\lambda$. We also note that our result here is true
for general `d' dimensional reduction and in order to obtain the O(d, d) 
invariance in that case, $B_{\mu\nu}^{(1)}$ should transform as,
\begineq
\tilde{B}_{\mu\nu}^{(1)} = B_{\mu\nu}^{(1)} - {\cal A}_\mu^T \epsilon
{\cal A}_\nu
\endeq
where $\epsilon = \left(\begin{array}{cc}
0  &  I \\
-I   &  0\end{array}\right)$ and ${\cal A}_\mu = \left(\begin{array}{c}
A_\mu^{(2)\,m}\\A_{\mu\,m}^{(3)}
\end{array}\right)$ with $I$ representing the d-dimensional identity matrix
and $m=1,2,\ldots,d$.
It is clear that the full action (42) does not have the global non-compact 
O(1, 1) symmetry under (46). We will, however, 
come back to this question later in section III. 

Finally, we point out as also noted in ref.[7], that the whole action (42) has
an obvious  global O(2) invariance since it is obtained from an eleven 
dimensional theory. In the process of dimensional reduction the original
Lorentz group SO(1, 10) gets split up into SO(1, 8) $\times$ O(2). So, the
resulting nine dimensional theory inherits a global O(2) invariance. Indeed
one can check that the action (42) is invariant under the following O(2)
transformations:
\begineq
\begin{array}{l}
\delta a \,\,\,=\,\,\, -\theta
\left(1+a^2-\chi^{\frac{1}{2}} e^{-2\phi}\right)\\
\delta e^{\phi} \,\,\,=\,\,\, \frac{7}{4} \theta a e^{\phi}\\
\delta \chi \,\,\,=\,\,\, -\theta \chi a\\
\delta A_\mu^{(3)} \,\,\,=\,\,\, 0\\
\delta A_\mu^{(i)} \,\,\,=\,\,\, \theta \epsilon^{ij} A_\mu^{(j)}\end{array}
\qquad\qquad
\begin{array}{l}
\delta g_{\mu\nu} \,\,\,=\,\,\, \theta g_{\mu\nu} a\\
\delta \bar{B}_{\mu\nu}^{(i)} \,\,\,=\,\,\, 
\theta \epsilon^{ij} \bar{B}_{\mu\nu}^{(j)}\\
\delta H_{\mu\nu\lambda}^{(i)} \,\,\,=\,\,\, 
\theta \epsilon^{ij} H_{\mu\nu\lambda}^{(j)}\\
\delta C_{\mu\nu\lambda} \,\,\,=\,\,\, \delta \bar{C}_{\mu\nu\lambda} = 0\\
\delta F_{\mu\nu\lambda\rho} \,\,\,=\,\,\, 0\end{array}
\endeq
where $\theta$ is a constant infinitesimal parameter. In the above the indices
$i$, $j$ = 1, 2 with $\epsilon^{12} = - \epsilon^{21} = 1$. Also, we note that
although the string frame metric transforms under the O(2) transformation, 
the Einstein frame metric in nine
dimensions $\bar{g}_{\mu\nu} = e^{-\frac{4}{7} \phi} g_{\mu\nu}$ does not.
We have written down our definitions of
various fields with explicit O(2) indices `(1)' and `(2)'. It is just a 
straightforward exercise to verify the invariance of the action (42) under
the transformations (50). Note that the topological term and the $F^2$ term
are individually invariant under the O(2) transformation.
It should be pointed out that under the O(2) transformation it is
$\bar{B}_{\mu\nu}^{(i)}$ which has a nice transformation property and not
$B_{\mu\nu}^{(i)}$. Also it is much 
easier to verify the invariance of the action (42) in the Einstein frame since
the Einstein metric remains inert under the O(2) transformation. 
Many interesting consequences with a finite version 
of this O(2) symmetry has been pointed out in ref.[7]. 

\vspace{1cm}

\begin{large}
\noindent{\bf III. Reduction of Type IIB Theory on $S^1$ and 
Duality Symmetries:}
\end{large}

\vspace{.5cm}

It is well-known that the equations of motion of the $N=2$, $D=10$ chiral
supergravity theory can not be obtained from a covariant action in ten 
dimensions. The bosonic sector of this theory contains a four-form gauge
potential whose field-strength is self-dual. If one sets this five-form field
strength to zero, then it is known that the equations of motion can be 
derived from a covariant action. We, therefore, consider this action
which is the low energy effective action of the type IIB string theory.
The action has the following form:
\eqabegin
S_{{\rm IIB}}^{(10)} &=&
\int\,d^{10}\h x \sqrt {-\hat {g}_B} \left[ e^{-2\hat{\phi}_B}
\left(\hat{R}_B + 4
\partial_{\h \mu} \hat{\phi}_B\partial^{\h \mu} \hat{\phi}_B - 
{1\over 12} \hat{h}_
{\h \mu \h \nu \h \lambda}^{(1)}\hat{h}^{(1)\,\h \mu \h \nu \h \lambda}
\right)\right.\nn\\
& &\qquad \left. -{1\over 2}\partial_{\h \mu} \hat{\psi}\partial^{\h \mu} 
\hat{\psi}
-{1\over 12}\left(\hat{h}_{\h \mu \h \nu \h \lambda}^{(2)} + 
\hat{\psi} \hat{h}_
{\h \mu \h \nu \h \lambda}^{(1)}\right)\left(\hat{h}^{(2)\,\h \mu \h \nu
\h \lambda} +
\hat{\psi} \hat{h}^{(1)\,\h \mu \h \nu \h \lambda}\right)\right]
\eqaend
We have denoted the metric, the scalar curvature and the dilaton with a
subscript `$B$' and the field 
strength with small letters to distinguish them from the corresponding 
fields in the type 
IIA theory. The type IIB theory (as also common to all other 
string theories) is
known to contain a metric, a dilaton and a two-form gauge field $\h b_{\h \mu
\h \nu}^{(1)}$ (with field strength $\hat{h}^{(1)\,\h \mu \h \nu \h \lambda}$)
in the NS-NS sector which is represented by the first three terms in (51)
with the usual dilaton coupling. The R-R sector consists of a scalar
$\h \psi$, a two-form gauge field $\h b_{\h \mu
\h \nu}^{(2)}$ (with field strength $\hat{h}^{(2)\,\h \mu \h \nu \h \lambda}$)  
and a four-form gauge field whose field strength has been set to zero. 
The R-R sector does
not couple to the dilaton as is clear from the last two terms in (51) and
contains the non-perturbative information of type IIB string theory. It is 
known that the action (51) possesses a global SL(2, R) invariance [20, 21]. 
This 
symmetry can be better understood in the Einstein frame since the 
Einstein metric remains inert under the SL(2, R) transformation. So, in order
to show this symmetry we first express the action in the Einstein metric
$\h {\bar{g}}_{B, \h \mu \h \nu} = e^{-\frac{1}{2} \h \phi} \h g_{B, \h \mu
\h \nu}$ as follows:
\eqabegin
\bar{S}_{{\rm IIB}}^{(10)} &=&
\int\,d^{10}\h x \sqrt {-\bar{\hat {g}}_B} \left[
\bar{\hat{R}}_B - \frac{1}{2}
\partial_{\h \mu} \hat{\phi}_B\partial^{\h \mu} \hat{\phi}_B -
\frac{1}{2}e^{2\h \phi_B} \partial_{\h \mu} \hat{\psi}\partial^{\h \mu}
\hat{\psi}\right.\nn\\
& & \left.-\frac{1}{12} \left(e^{-\h \phi_B}\hat{h}_
{\h \mu \h \nu \h \lambda}^{(1)}\hat{h}^{(1)\,\h \mu \h \nu \h \lambda}
+ e^{\h \phi_B}\left(\hat{h}_{\h \mu \h \nu \h \lambda}^{(2)} +
\hat{\psi} \hat{h}_
{\h \mu \h \nu \h \lambda}^{(1)}\right)\left(\hat{h}^{(2)\,\h \mu \h \nu
\h \lambda} +
\hat{\psi} \hat{h}^{(1)\,\h \mu \h \nu \h \lambda}\right)\right)\right]
\eqaend
This action can now be expressed in a manifestly SL(2, R) invariant form
as given below,
\begineq
\bar{S}_{{\rm IIB}}^{(10)}\,\,=\,\,
\int\,d^{10}\h x \sqrt {-\bar{\hat {g}}_B} \left[
\bar{\hat{R}}_B + \frac{1}{4}{\rm tr}
\partial_{\h \mu}S^{-1}\partial^{\h \mu}S-
\frac{1}{12}\h {\cal H}^T_{\h \mu \h \nu \h \lambda}
S \h {\cal H}^{\h \mu \h \nu \h \lambda}\right]
\endeq
where 
\begineq
S = \left(\begin{array}{cc}
\h \psi^2 e^{\h \phi_B} + e^{-\h \phi_B}  & \h \psi e^{\h \phi_B}  \\
      \h \psi e^{\h \phi_B}            &  e^{\h \phi_B}\end{array}\right)
\endeq
represents an SL(2, R) matrix and $\h {\cal H} = \left(\begin{array}{c}
\h h_{\h \mu \h \nu \h \lambda}^{(1)}  \\
\h h_{\h \mu \h \nu \h \lambda}^{(2)}\end{array}\right)$. Under a global 
SL(2, R) transformation $S\rightarrow \Lambda S \Lambda^T$ and 
$\left(\begin{array}{c}
\h b_{\h \mu \h \nu}^{(1)}  \\
\h b_{\h \mu \h \nu}^{(2)}\end{array}\right) = \h {\cal B}_{\h \mu \h \nu}
\rightarrow (\Lambda^{-1})^T \h {\cal B}_{\h \mu\h \nu}$, the action (53) is 
easily seen to be invariant. With these transformations the complex scalar
field $\h \rho = \left(\h \psi +i e^{-\h \phi_B}\right)$ undergoes a 
fractional linear transformation whereas the two two-form potentials
transform linearly. In particular, choosing $\Lambda = 
\left(\begin{array}{cc}
 0 & 1  \\
-1 & 0\end{array}\right)$ and $\h \psi = 0$, the string coupling constant
transforms to its inverse showing a strong-weak coupling duality in the
theory. We will point out later that this symmetry has its origin in the
eleven dimensional CJS theory  compactifying  on a torus $T^2$.

Next, we dimensionally reduce the action (51) on $S^1$. Since we have studied 
the dimensional reduction of type IIA theory in detail in section II, we
here give the results. The reduced form of the action is given below:
\eqabegin
S^{(9)}_{{\rm IIB}} &=&
\int\,d^{9}x\sqrt {-g_B} \left[ e^{-2\phi_B}\left(R_B + 4
\partial_\mu \phi_B\partial^\mu \phi_B -\frac{1}{4} \partial_\mu \log \chi_B
\partial^\mu \log \chi_B  \right.\right.\nn\\
& &\qquad\qquad\left.-\frac{1}{4} \chi_B
f_{\mu\nu}^{(3)} f^{(3)\,\mu\nu}
- \frac{1}{4}\chi_B^{-1} f_{\mu\nu}^{(1)}
f^{(1)\,\mu\nu} -\frac{1}{12} h_
{\mu\nu\lambda}^{(1)} h^{(1)\,\mu\nu\lambda}
\right) -\frac{1}{2} \chi_B^{\frac{1}{2}}\partial_\mu \psi
\partial^\mu \psi\nn\\
& &\qquad\qquad -\frac{1}{12}\chi_B^{\frac{1}{2}}
\left(h_{\mu\nu\lambda}^{(2)} +\psi h_{\mu\nu
\lambda}^{(1)}\right)\left(h^{(2)\,\mu\nu\lambda}
+\psi h^{(1)\,\mu\nu\lambda}
\right)\nn\\
& &\qquad\qquad\left. -\frac{1}{4} \chi_B^{-\frac{1}{2}}
\left(f_{\mu\nu}^{(2)} + \psi f_{\mu\nu}^{(1)}
\right)\left(f^{(2)\,\mu\nu} + \psi f^{(1)\,\mu\nu}\right)\right]
\eqaend
where our definitions of fields and the corresponding field strengths are:
\eqabegin
\h g_{B \h \mu\h \nu}&\longrightarrow & 
\left\{ \begin{array}{l} \h g_{B,99} = g_{B,99} = \chi_B
\\
\h g_{B \mu 9} = g_{B,\mu 9} = \chi_B a_\mu^{(3)} \\
\h g_{B \mu\nu}= g_{B \mu\nu} + \chi_B a_\mu^{(3)} a_\nu^{(3)} 
\end{array}\right.\\
\h \phi_B &=& \phi_B + \frac{1}{4} \log \chi_B\\
\h b_{\h \mu \h \nu}^{(1)} & \longrightarrow & \left\{ \begin{array}{l}
b_{\mu 9}^{(1)} = \h b_{\mu 9}^{(1)} = a_\mu^{(1)} \\
b_{\mu\nu}^{(1)}
= \h b_{\mu\nu}^{(1)} + a_\mu^{(3)} a_{\nu}^{(1)} - a_\nu^{(3)} a_{\mu}^{(1)}
\end{array}\right.\\
\h b_{\h \mu \h \nu}^{(2)} &\longrightarrow & \left\{ \begin{array}{l}
b_{\mu 9}^{(2)} = \h b_{\mu 9}^{(2)} = a_\mu^{(2)} \\
b_{\mu\nu}^{(2)}
= \h b_{\mu\nu}^{(2)} + a_{\mu}^{(3)} a_{\nu}^{(2)} -
a_{\nu}^{(3)} a_{\mu}^{(2)}
\end{array}\right.\\
\h \psi &=& \psi
\eqaend
The corresponding field strengths are:
\eqabegin
f_{\mu\nu}^{(3)} &=& \partial_\mu a_\nu^{(3)} - \partial_\nu a_\mu^{(3)}\\
h_{\mu\nu 9}^{(1)} &=& \h h_{\mu\nu 9}^{(1)}\,\,\,=\,\,\,f_{\mu\nu}^{(1)}
\,\,\,=\,\,\, \partial_\mu a_\nu^{(1)} - \partial_\nu a_\mu^{(1)}\\
h_{\mu\nu \lambda}^{(1)} &=& \partial_\mu b_{\nu\lambda}^{(1)} -
f_{\mu\nu}^{(3)} a_\lambda^{(1)} + {\rm cyc.\,\, in}\,\,\,\mu\nu\lambda\nn\\
&=& \partial_\mu \bar{b}_{\nu\lambda}^{(1)} -\frac{1}{2}\left(
f_{\mu\nu}^{(3)} a_\lambda^{(1)} + f_{\mu\nu}^{(1)} a_\lambda^{(3)}\right)
+ {\rm cyc.\,\, in}\,\,\,\mu\nu\lambda
\eqaend
where we have defined
\begineq
\bar{b}_{\mu\nu}^{(1)} \equiv b_{\mu\nu}^{(1)} - \frac{1}{2}\left(
a_{\mu}^{(3)} a_\nu^{(1)} - a_{\nu}^{(3)} a_\mu^{(1)}\right)
\endeq
and finally,
\eqabegin
h_{\mu\nu 9}^{(2)} &=& \h h_{\mu\nu 9}^{(2)}\,\,\,=\,\,\,f_{\mu\nu}^{(2)}
\,\,\,=\,\,\, \partial_\mu a_\nu^{(2)} - \partial_\nu a_\mu^{(2)}\\
h_{\mu\nu \lambda}^{(2)} &=& \partial_\mu b_{\nu\lambda}^{(2)} -
f_{\mu\nu}^{(3)} a_\lambda^{(2)} + {\rm cyc.\,\, in}\,\,\,\mu\nu\lambda\nn\\
&=& \partial_\mu \bar{b}_{\nu\lambda}^{(2)} -\frac{1}{2}\left(
f_{\mu\nu}^{(3)} a_\lambda^{(2)} + f_{\mu\nu}^{(2)} a_\lambda^{(3)}\right)
+ {\rm cyc.\,\, in}\,\,\,\mu\nu\lambda
\eqaend
with
\begineq
\bar{b}_{\mu\nu}^{(2)} \equiv b_{\mu\nu}^{(2)} -\frac{1}{2}\left(
a_{\mu}^{(3)} a_\nu^{(2)} - a_{\nu}^{(3)} a_\mu^{(2)}\right)
\endeq
We again notice here that the NS-NS sector of the reduced action (55) 
represented by the first six terms has a non-compact global O(1, 1) symmetry
of the form $M\rightarrow \Omega M \Omega^T$ and ${\cal A}_\mu
\rightarrow \Omega {\cal A}_\mu$ where,
\begineq
M = \left(\begin{array}{cc}
\chi_B^{-1}  &  0 \\
       0   &  \chi_B\end{array}\right)
\qquad {\rm and}\qquad
{\cal A}_{\mu} = \left(\begin{array}{c}
a_{\mu}^{(3)}\\ a_{\mu}^{(1)}
\end{array}\right)
\endeq
Note that $M$ is an O(1, 1) matrix satisfying $M^T \eta M = \eta$ with
$\eta = \left(\begin{array}{cc}
0  &  1 \\
       1   &  0\end{array}\right)$. $\Omega$ is the O(1, 1) matrix of
transformation which has the form $\left(\begin{array}{cc}
0  &  \lambda \\
       \lambda^{-1}   &  0\end{array}\right)$. Under the O(1, 1)
transformation, the moduli field $\chi_B$ and the vector gauge fields 
transform as
\eqabegin
\tilde {\chi}_B &=& \lambda^{-2} \chi^{-1}_B\nn\\
\tilde {a}_\mu^{(3)} &=& \lambda a_\mu^{(1)}\\
\tilde {a}_\mu^{(1)} &=& \lambda^{-1} a_\mu^{(3)}\nn
\eqaend 
Also, note that in order to recover the O(1, 1) invariance of the NS-NS
sector $\bar{b}_{\mu\nu}^{(1)}$ does not transform whereas 
$b_{\mu\nu}^{(1)}$ transforms as,
\begineq
\tilde{b}_{\mu\nu}^{(1)} = b_{\mu\nu}^{(1)} - \left(a_\mu^{(3)}a_\nu^{(1)}
- a_\nu^{(3)}a_\mu^{(1)}\right)
\endeq
The metric $g_{B,\mu\nu}$ and the dilaton $\phi_B$ do not transform under the
O(1, 1) transformation.

As noted before for the type IIA theory, we again notice that the full action
(55) does not remain invariant under the O(1, 1) transformation but changes to
\eqabegin
\tilde{S}^{(9)}_{{\rm IIB}} &=&
\int\,d^{9}x\sqrt {-g_B} \left[ e^{-2\phi_B}\left(R_B + 4
\partial_\mu \phi_B\partial^\mu \phi_B -\frac{1}{4} \partial_\mu \log \chi_B
\partial^\mu \log \chi_B  \right.\right.\nn\\
& &\qquad\qquad\left.-\frac{1}{4} \chi_B
f_{\mu\nu}^{(3)} f^{(3)\,\mu\nu}
- \frac{1}{4}\chi_B^{-1} f_{\mu\nu}^{(1)}
f^{(1)\,\mu\nu} -\frac{1}{12} h_
{\mu\nu\lambda}^{(1)} h^{(1)\,\mu\nu\lambda}
\right)\nn\\ 
& &\qquad\qquad -\frac{1}{2}\lambda^{-1} 
\chi_B^{-\frac{1}{2}}\partial_\mu \psi
\partial^\mu \psi
-\frac{1}{12}\lambda^{-1}\chi_B^{-\frac{1}{2}}
\left(h_{\mu\nu\lambda}^{(2)} +\psi h_{\mu\nu
\lambda}^{(1)}\right)\left(h^{(2)\,\mu\nu\lambda}
+\psi h^{(1)\,\mu\nu\lambda}
\right)\nn\\
& &\qquad\qquad\left. -\frac{1}{4} \lambda\chi_B^{\frac{1}{2}}
\left(f_{\mu\nu}^{(2)} + \lambda^{-1}\psi f_{\mu\nu}^{(3)}
\right)\left(f^{(2)\,\mu\nu} + \lambda^{-1}\psi f^{(3)\,\mu\nu}\right)\right]
\eqaend
We have assumed here, that like $h_{\mu\nu\lambda}^{(1)}$, the other
field strength $h_{\mu\nu\lambda}^{(2)}$ also does not transform under
under O(1, 1) transformation. This, however, means that the R-R two-form
potential $b_{\mu\nu}^{(2)}$ (also $\bar {b}_{\mu\nu}^{(2)}$) transforms
in a non-trivial non-local way under the duality transformation. Also,
the scalar $\psi$ is assumed to remain inert under the O(1, 1)
transformation. 
Now comparing this T-dual action (71) and the nine dimensional type IIA 
action (42),
we find that they precisely match if the fields in the type IIA theory
satisfy the following relations:
\eqabegin
F_{\mu\nu\rho\lambda} & = & 0\\
\epsilon^{\mu_1\ldots \mu_9} \epsilon^{ij} H_{\mu_1\mu_2\mu_3}^{(i)}
H_{\mu_4\mu_5\mu_6}^{(j)}\left(\bar{C}_{\mu_7\mu_8\mu_9} + 3 A_{\mu_7}^{(k)}
\bar{B}_{\mu_8\mu_9}^{(k)}\right) & = & 0
\eqaend 
and we make the following field identifications:
\begineq
\begin{array}{l}
g_{B,\mu\nu}\,\,\,\equiv\,\,\,g_{\mu\nu}\\
\phi_B\,\,\,\equiv\,\,\,\phi\\
\lambda^2 \chi_B\,\,\,\equiv\,\,\,\chi\\
\psi\,\,\,\equiv\,\,\,-a\\
\lambda^{-1} a_\mu^{(3)}\,\,\,\equiv\,\,\,-A_{\mu}^{(2)}
\end{array}
\qquad\qquad\qquad
\begin{array}{l}
\lambda a_\mu^{(1)}\,\,\,\equiv\,\,\,A_{\mu}^{(3)}\\
a_\mu^{(2)}\,\,\,\equiv\,\,\,A_\mu^{(1)}\\
h_{\mu\nu\lambda}^{(1)}\,\,\,\equiv\,\,\,H_{\mu\nu\lambda}^{(1)}\\
h_{\mu\nu\lambda}^{(2)}\,\,\,\equiv\,\,\,H_{\mu\nu\lambda}^{(2)}
\end{array}
\endeq
Here eq.(72) simply says that the dimensionally reduced four form 
field strength in type IIA theory has to be zero, whereas eq.(73) says
that the dual of the combination $\bar{C}_{\mu\nu\lambda} 
+ \left(A_{\mu}^{(i)}\bar{B}_
{\nu\lambda}^{(i)} + {\rm cyc.\,\, in\,\,\, \mu\nu\lambda}\right)$
is transverse to the $H$'s.
Note also that the identification (74) is consistent in the sense that 
the gauge fields
$a_\mu^{(3)}$ and $A_\mu^{(2)}$ appear from the dimensional reduction of the
ten dimensional metric in type IIB theory and IIA theory respectively. Also
$a_\mu^{(1)}$ and $A_\mu^{(3)}$ appear from the reduction of the two-form 
potential in ten dimensions. The other fields in these two theories have
completely different origin and so, there is no contradiction in identifying 
them in nine dimensions. We would like to point out here that both the
antisymmetric two-form potentials (NS-NS and R-R) in the two theories are 
related to each other in non-local way. We have thus shown, 
purely from the bosonic consideration,
that type IIA and type IIB theories are T-dual to each other.

We now show that the O(1, 1) or the T-duality transformation (69) is 
nothing but
the Buscher's duality transformation [25] of the various components of 
the metric,
antisymmetric tensor field and the dilaton in ten dimensions for a particular
value of $\lambda$. Since the following relations are valid for the common
NS-NS sector of any string theory we will omit the index `$B$'. The various 
components of the metric, the antisymmetric tensor field and the dilaton
in ten and nine dimensions can be seen from eqs.(56), (58) and (57) to be 
related as,
\eqabegin
\h g_{99} &=& g_{99}\,\,\,=\,\,\,\chi\nn\\
\h g_{\mu 9} &=& g_{99} a_\mu^{(3)}\,\,\,=\,\,\,\chi a_\mu^{(3)}\nn\\
\h g_{\mu\nu} &=& g_{\mu\nu} + \chi a_\mu^{(3)} a_\nu^{(3)}\nn\\
\h b_{\mu 9}^{(1)} &=& b_{\mu 9}^{(1)}\,\,\,=\,\,\,a_\mu^{(1)}\\
\h b_{\mu \nu}^{(1)} &=& \bar{b}_{\mu \nu}^{(1)} - \frac{1}{2}\left(
a_\mu^{(3)} a_\nu^{(1)} - a_\nu^{(3)} a_\mu^{(1)}\right)\nn\\
\h \phi &=& \phi + \frac{1}{4} \log \chi\,\,\,=\,\,\, 
\phi + \frac{1}{4} \log g_{99}\nn
\eqaend
We have expressed $\h b_{\mu \nu}^{(1)}$ in terms of $\bar{b}_{\mu \nu}^{(1)}$
because we know that it does not transform under O(1, 1) transformation (69).
This is crucial to obtain the Buscher's duality rule [22] correctly. Under the
duality transformation, they transform to
\eqabegin
\tilde{\h g}_{99} &=& \lambda^{-2}\chi^{-1}\,\,\,=\,\,\,\frac{1}{\lambda^2
\h g_{99}}\\
\tilde{\h g}_{\mu 9} &=& \lambda^{-1}\chi^{-1} a_\mu^{(1)}\,\,\,=\,\,\,
\frac{\h b_{\mu 9}^{(1)}}{\lambda \h g_{99}}\\
\tilde{\h g}_{\mu\nu} &=& \h g_{\mu\nu} - \chi a_\mu^{(3)} a_\nu^{(3)}
+\chi^{-1}  a_\mu^{(1)} a_\nu^{(1)}\nn\\
&=& \h g_{\mu\nu} -\frac{1}{\h g_{99}}\left(\h g_{\mu 9} \h g_{\nu 9}
-\h b_{\mu 9}^{(1)} \h b_{\nu 9}^{(1)}\right)\\
\tilde{\h b}_{\mu 9}^{(1)} &=& \lambda^{-1}a_{\mu}^{(3)}\,\,\,=\,\,\,
\frac{1}{\lambda}\frac{\h g_{\mu 9}}{\h g_{99}}\\
\tilde{\h b}_{\mu \nu}^{(1)} &=& \h b_{\mu \nu}^{(1)} + \left(
a_\mu^{(3)} a_\nu^{(1)} - a_\nu^{(3)} a_\mu^{(1)}\right)\nn\\
&=& \h b_{\mu \nu}^{(1)} + \frac{1}{\h g_{99}}\left(
\h g_{\mu 9} \h b_{\nu 9}^{(1)} - \h g_{\nu 9} \h b_{\mu 9}^{(1)}\right)\\
\tilde{\h \phi} &=& \h \phi - \frac{1}{2} \log \chi - \frac{1}{2} 
\log|\lambda|\,\,\,=\,\,\,\h \phi -\frac{1}{2} \log \h g_{99} - \frac{1}{2}
\log|\lambda|
\eqaend
We point out that these O(1, 1) transformations (76)--(81) match precisely with 
Buscher's rule for $\lambda = -1$. For generic $\lambda$, however, we
do not get any new information from (76)--(81) since it is clear that
$\lambda$ can be scaled away in all the expressions except the dilaton
transformation rule in (81) by scaling the ninth coordinate $x^9 \rightarrow
\lambda x^9$. In (81), this scaling does not completely absorb the $\lambda$
term, but, since $\lambda$ is a constant, gives a constant shift in the 
dilaton which can be absorbed into the gravitational constant in front of the 
string effective action not displayed explicitly.

We would also like to comment that since we have mapped the type IIB theory
in nine dimensions exactly to the type IIA theory by an O(1, 1) 
or T-duality transformation by field identifications, the type IIB theory also
possesses a global O(2) invariance in nine dimensions very much like the 
type IIA
theory. The complete O(2) transformations in the type IIB theory are given by
\begineq
\begin{array}{l}
\delta \psi \,\,\,=\,\,\, \theta_B
\left(1+\psi^2-\chi_B^{-\frac{1}{2}} e^{-2\phi_B}\right)\\
\delta e^{\phi_B} \,\,\,=\,\,\, -\frac{7}{4} \theta_B \psi e^{\phi_B}\\
\delta \chi_B \,\,\,=\,\,\, -\theta_B \chi_B \psi\\
\delta a_\mu^{(3)} \,\,\,=\,\,\, 0\\
\end{array}
\qquad\qquad
\begin{array}{l}
\delta a_\mu^{(i)} \,\,\,=\,\,\, \theta_B \epsilon^{ij} a_\mu^{(j)}\\
\delta g_{B,\mu\nu} \,\,\,=\,\,\, -\theta_B g_{B,\mu\nu} \psi\\
\delta \bar{b}_{\mu\nu}^{(i)} \,\,\,=\,\,\,
\theta_B \epsilon^{ij} \bar{b}_{\mu\nu}^{(j)}\\
\delta h_{\mu\nu\lambda}^{(i)} \,\,\,=\,\,\,
\theta_B \epsilon^{ij} h_{\mu\nu\lambda}^{(j)}
\end{array}
\endeq
Here $\theta_B$ is a constant infinitesimal parameter and $i$, $j$ = 1, 2 with
$\epsilon^{12} = -\epsilon^{21} = 1$. One can check indeed that the action (55)
is invariant under this transformation although type IIB theory is not 
obtained from the dimensional reduction of an eleven dimensional theory.

It is quite natural to expect that the type IIB theory in nine dimensions (55)
should have a manifest global SL(2, R) invariance like its parent theory in 
ten dimensions. We say it is natural, because the matrix $S$ eq.(54) 
that was constructed to show the 
SL(2, R) invariance of the action only involves the 
scalar fields which remain intact under
dimensional reduction. We, however, show that this expectation fails primarily
because the matrix S in (54) is manifestly dependent on the number of the
space-time dimensions.
To show this it is enough to consider a simpler case when $\chi_B$ = 1, i.e. 
at the 
self-dual point. (The general case can also be worked out quite easily.) 
We first rewrite the action (55) in the Einstein frame since
the Einstein metric remains inert under SL(2, R) transformation. With the 
Weyl scaling $\bar{g}_{B,\mu\nu} = e^{-\frac{4}{7}\phi_B} g_{B,\mu\nu}$ the
action takes the form: 
\eqabegin
\bar{S}^{(9)}_{{\rm IIB}} &=&
\int\,d^{9}x\sqrt {-\bar{g}_B} \left[\bar {R}_B -\frac{4}{7}
\partial_\mu \phi_B\partial^\mu \phi_B -\frac{1}{2} 
e^{2\phi_B}\partial_\mu \psi
\partial^\mu \psi \right.\nn\\
& &\qquad\qquad -\frac{1}{4} e^{-\frac{4}{7}\phi_B}\left(
f_{\mu\nu}^{(3)} f^{(3)\,\mu\nu}
+  f_{\mu\nu}^{(1)}
f^{(1)\,\mu\nu}\right) -\frac{1}{12} e^{-\frac{8}{7}\phi_B} h_
{\mu\nu\lambda}^{(1)} h^{(1)\,\mu\nu\lambda}\nn\\
& &\qquad\qquad -\frac{1}{12} e^{\frac{6}{7}\phi_B}
\left(h_{\mu\nu\lambda}^{(2)} +\psi h_{\mu\nu
\lambda}^{(1)}\right)\left(h^{(2)\,\mu\nu\lambda}
+\psi h^{(1)\,\mu\nu\lambda}
\right)\nn\\
& &\qquad\qquad\left. -\frac{1}{4} e^{\frac{10}{7}\phi_B}
\left(f_{\mu\nu}^{(2)} + \psi f_{\mu\nu}^{(1)}
\right)\left(f^{(2)\,\mu\nu} + \psi f^{(1)\,\mu\nu}\right)\right]
\eqaend
We note that in order to reproduce the dilaton as well as the scalar term
in the action correctly in any dimension, we should choose the matrix $S$
to have the following form: (It is worth emphasizing here that the dependence
on D, alternately, could be transferred to the exponential which is equivalent
to rescaling the dilaton. In fact, this is the way one gets 
SL(2, R) invariance in four dimensional string theory.)
\begineq
S \,\,\,=\,\,\, e^{\phi_B}\left( \begin{array}{cc}
\left(\frac{D-2}{8}\right) \psi^2 + e^{-2\phi_B} & \psi 
\left(\frac{D-2}{8}\right)^{\frac{1}{2}}\\
\psi \left(\frac{D-2}{8}\right)^{\frac{1}{2}} & 1
\end{array}\right)
\endeq
so that
\begineq
\frac{2}{D-2}{\rm tr}\partial_\mu S^{-1}
\partial^\mu S\,\,\,=\,\,\ -\frac{4}{D-2} \partial_\mu \phi_B
\partial^\mu \phi_B - \frac{1}{2} e^{2\phi_B} \partial_\mu \psi\partial^\mu
\psi
\endeq
In particular, in ten dimensions (85) correctly reproduces the scalar terms
and it matches with the form of $S$ given in eq.(54). Also, for nine dimensions
it produces the scalar terms in (83) correctly. But it is the explicit 
$D$-dependence in the matrix $S$ which spoils the manifest SL(2, R)
invariance of the other terms in the action. For example, if we want to
produce $h_{\mu\nu\lambda}$ terms we choose
\begineq
{\cal H}_{\mu\nu\lambda}\,\,\,=\,\,\,\left( \begin{array}{c}
e^{-\frac{1}{14}\phi_B} h_{\mu\nu\lambda}^{(1)} \\
e^{-\frac{1}{14}\phi_B} h_{\mu\nu\lambda}^{(2)}\end{array}\right)
\endeq
then,
\eqabegin
{\cal H}_{\mu\nu\lambda}^T S {\cal H}^{\mu\nu\lambda} &=& 
e^{-\frac{8}{7}\phi_B} h_{\mu\nu\lambda}^{(1)}h^{(1)\,\mu\nu\lambda}\nn\\
& & + e^{\frac{6}{7}\phi_B}\left(h_{\mu\nu\lambda}^{(2)} + \psi \left(
\frac{D-2}{8}\right)^{\frac{1}{2}} h_{\mu\nu\lambda}^{(1)}\right)
\left(h^{(2)\,\mu\nu\lambda} + \psi \left(
\frac{D-2}{8}\right)^{\frac{1}{2}} h^{(1)\,\mu\nu\lambda}\right)\nn\\
\eqaend
So, for $D$=9 we almost get the $ h_{\mu\nu\lambda}$ term in (83) except
the factor $\left(\frac{D-2}{8}\right)^{\frac{1}{2}}$. Note that 
the exponential 
factor can always be adjusted properly by scaling the other fields in an
appropriate way. For $D$=10, it produces the $h_{\mu\nu\lambda}$ 
term correctly
because in that case $\frac{D-2}{8} = 1$. But for other $D$ we will not get
$ h_{\mu\nu\lambda}$ term correctly since the factor can not be scaled away.
Thus, we conclude that eventhough the ten dimensional type IIB theory has
a manifest SL(2, R) invariance, this manifest symmetry gets spoiled by
dimensional reduction.

It can also be checked in a similar fashion that the type IIA string
effective action in nine dimensions, eq.(42), does not have a manifest
SL(2, R) invariance even when the conditions (72) and (73) are satisfied. 
It is
also not expected since the parent theory eq.(17) does not have this
symmetry in any manifest way at least in terms of the variables in which the
action is written. We, however, give here a dynamical argument to show
that like in type IIB theory in ten dimensions, type IIA theory also
should have an SL(2, R) S-duality invariance in ten dimensions.

It has been observed by Aspinwall [24] that the strong-weak coupling duality
(which we have interchangably called an SL(2, R) S-duality) symmetry
in type IIB string theory in ten dimensions has its origin in the
underlying eleven dimensional theory by compactifying it on $S^1 \times
S^1 \cong T^2$. Since we have also studied the dimensional reduction
of eleven dimensional CJS theory, we can reformulate his argument in 
terms of the fields we have defined in section II. Note from (15), that
when we compactified the CJS theory on $S^1$ to obtain type IIA theory,
the radius of the circle of compactification was found to be related
to the ten dimensional dilaton by
\begineq
R_1 \sim e^{\frac{2}{3}\h \phi}
\endeq
So, the type IIA string coupling constant is given by
\begineq
\lambda_{{\rm A}}\,\,=\,\,e^{\h \phi} \sim R_1^{\frac{3}{2}}
\endeq
When this type IIA theory was further compactified on second $S^1$, the radius
of the circle was chosen to be $\chi^{\frac{1}{2}}$ (see eq.(25)). This
radius is mesured in string metric. Because, we had to rescale the ten
dimensional metric (see eq. (14)) to obtain type IIA theory, so the radius
of the second circle in terms of the original metric would be given as
\begineq
R_2 \sim e^{-\frac{1}{3}\h \phi} \chi^{\frac{1}{2}}
\endeq
So, writing in terms of the nine dimensional fields we get
\eqabegin
R_1 &\sim & e^{\frac{2}{3}\phi} \chi^{\frac{1}{6}}\nn\\
R_2 &\sim & e^{-\frac{1}{3}\phi} \chi^{\frac{5}{12}}
\eqaend
Since the string coupling constant of the type IIA theory $\lambda_{{\rm A}}
= e^{(\phi +\frac{1}{4} \log \chi)}$ changes to that of type IIB theory
under the T-duality transformation $\tilde {\chi} = \chi^{-1}$ (we have
set $\lambda = -1$ as this produces the Buscher's duality transformation),
we have
\begineq
\lambda_{{\rm B}}\,\,=\,\,e^{(\phi -\frac{1}{4} \log \chi)} \sim 
\frac{R_1}{R_2}
\endeq
We point out that the nine dimensional type IIA theory should have a
symmetry if we interchange $R_1$ and $R_2$. In other words, we should have
the same theory if we compactify CJS theory first on $S^1$ with radius
$R_1$ and then on the other $S^1$ with radius $R_2$ or first on $S^1$ with
radius $R_2$ and then on the other $S^1$ with radius $R_1$. But this symmetry
of the nine dimensional theory has a dramatic consequence for the type IIB 
theory in ten dimensions since it takes the string coupling constant to
its inverse. Thus we find a strong-weak coupling duality symmetry in type IIB
theory as a consequence of $R_1\leftrightarrow R_2$ symmetry of type IIA theory
in nine dimensions. In fact, by taking into account the angle between $R_1$
and $R_2$, the SL(2, R) symmetry group of the type IIB theory can be identified
with the modular group of the torus $T^2 \cong S^1 \times S^1$ as also
observed in ref.[21]. 

We now compute the T-dual of $R_1$ and $R_2$ as given by,
\eqabegin
\tilde {R}_1 &\sim & e^{\frac{2}{3} \phi} \chi^{-\frac{1}{6}}\,\,\sim \,\,
\left(\frac{R_1}{R_2}\right)^{\frac{2}{3}}\,\,\sim \,\, \lambda_{{\rm B}}
^{\frac{2}{3}}\\
\tilde {R}_2 &\sim & e^{-\frac{1}{3} \phi} \chi^{-\frac{5}{12}}\,\,\sim \,\,
\left(\frac{R_1}{R_2}\right)^{\frac{2}{3}} R_1^{-\frac{3}{2}}
\,\,\sim \,\, \frac{\lambda_{{\rm B}}^{\frac{2}{3}}}{\lambda_{{\rm A}}}
\eqaend
Using (93) and (94) we find
\begineq
\lambda_{{\rm A}} \,\, \sim
\,\, \frac{\tilde {R}_1}{\tilde {R}_2}
\endeq
Since the nine dimensional theory should have an exchange symmetry $\tilde {R}
_1 \leftrightarrow \tilde {R}_2$, we thus find a strong-weak coupling duality
symmetry also in type IIA theory in ten dimensions. This symmetry has its 
origin in the eleven dimensional CJS theory by compactifying it on T-dual torus
$\tilde {T}^2$ with radii $\tilde {R}_1 = 
\left(\frac{R_1}{R_2}\right)^{\frac{2}{3}}$ and $\tilde {R}_2 = 
\left(\frac{R_1}{R_2}\right)^{\frac{2}{3}} R_1^{-\frac{3}{2}}$.
It remains an interesting puzzle to obtain this symmetry of the type IIA action
directly in ten dimensions.

\vspace{1 cm}

\begin{large}
\noindent{\bf IV. Conclusions:}
\end{large}

\vspace{.5cm}

We have performed the ``ordinary'' Scherk-Schwarz dimensional reduction of the
bosonic sector of the low energy effective action (CJS theory) of a 
hypothetical M-theory on $S^1 \times S^1 \cong T^2$. In this way we have 
obtained the low energy effective actions of type IIA string theory in both 
ten and nine space-time dimensions. Type IIA string effective actions obtained
this way not only contain the usual NS-NS gauge fields but also contain the
R-R gauge fields which do not couple to the dilaton and encode the 
non-perturbative information inherited from the eleven dimensional CJS theory.
Since the string coupling constant in ten dimensions is directly proportional 
to the radius of compactification of the circle, M-theory describes the
non-perturbative limit of type IIA string theory. The NS-NS sector of the nine
dimensional type IIA string effective action has a noncompact global O(1, 1)
invariance. We have pointed out how to recover this symmetry correctly under
which the moduli, the vector gauge fields as well as the antisymmetric
Kalb-Ramond field transform in a nontrivial way. The full action, including
the R-R sector in nine dimensions does not possess this symmetry, however,
it has a global O(2) invariance as a consequence of the Lorentz symmetry
of the original eleven dimensional theory. We then considered the type IIB
string effective action containing the R-R fields when the self-dual five-form
field strength is set to zero. This action is known to possess an SL(2, R)
duality invariance. We performed the dimensional reduction of this action 
on $S^1$. The NS-NS sector of the resulting nine dimensional theory also
has a noncompact global O(1, 1) invariance, but much like the type IIA theory
in nine dimensions, the full action does not have this symmetry. However, 
we have shown that under this O(1, 1) transformation the complete bosonic
sector of the type IIB theory
in nine dimensions reduces to the type IIA action with  proper field
identifications. In order to obtain this, the fields in the type IIA theory 
would have to satisfy the relations given in eqs.(72) and (73). 
This, therefore, shows  purely from 
the bosonic considerations that the type IIA and type 
IIB theories are T-dual to each
other. We then showed that for a particular value of the parameter of the
O(1, 1) transformation, it reduces precisely to the Buscher's duality rules
for the various components of the metric, the antisymmetric tensor field 
and the dilaton in ten dimensions. We have pointed out that the type IIB theory
in nine dimensions has a global O(2) invariance although it is not obtained
from an eleven dimensional theory. The original manifest SL(2, R) invariance
of the type IIB theory in ten dimensions have been shown to be spoiled by
the dimensional reduction and the nine dimensional theory does not have
this symmetry in any manifest way. We then presented a dynamical argument
to show that the type IIA theory in ten dimensions also has an SL(2, R)
S-duality invariance. This can be understood by compactifying CJS theory 
on T-dual torus $\tilde {T}^2$. It would be very interesting to realize 
this symmetry in a manifest way at the level of type IIA action in ten
dimensions.

\vspace{1cm}

\begin{large}

\noindent{\bf Acknowledgements:}

\end{large}

\vspace{.5cm}

We would like to thank E. Bergshoeff for discussions on some points. 
A.D. would like to thank the members of the Departamento de Fisica de
Particulas for hospitality and S.R. would like to thank A. V. Ramallo and 
J. M. Sanchez de Santos
for discussions. This work was supported in part by U.S.D.O.E. Grant No.
DE-FG-02-91ER40685. The work of S.R. has been supported in part by a 
fellowship from the Spanish Ministry of Education (MEC).

\vspace{1cm}

\begin{large}

\noindent{\bf References:}
\end{large}

\vspace{.5cm}

\begin{enumerate}
\item E. Witten, \np 443 (1995) 85.
\item E. Witten, {\it Some Comments On String Dynamics}, preprint 
IASSNS-HEP-63, hep-th/9507121. 
\item C. M. Hull, {\it String Dynamics at Strong Coupling}, preprint
QMW-95-50, hep-th/9512181.
\item E. Cremmer, B. Julia and J. Scherk, \pl 76 (1978), 409. 
\item E. Cremmer and B. Julia, \np 159 (1979) 141.
\item E. Witten as quoted in J. H. Schwarz, \pl 367 (1996) 97.
\item E. Bergshoeff, C. Hull and T. Ortin, \np 451 (1995) 547. 
\item J. Polchinski and E. Witten, {\it Evidence for Heterotic--Type I
String Duality}, preprint IASSNS-HEP-95-81, NSF-ITP-95-135, hep-th/9510169.
\item P. Horava and E. Witten, {\it Heterotic and Type I String Dynamics
from Eleven Dimensions}, preprint IASSNS-HEP-95-86, PUPT-1571, hep-th/9510209.  
\item J. Dai, R. G. Leigh and J. Polchinski, Mod. Phys. Lett. A4 (1989) 2073.
\item R. G. Leigh, Mod. Phys. Lett. A4 (1989) 2767. 
\item J. Polchinski, Phys. Rev. Lett. 75 (1995) 4724.
\item P. K. Townsend, {\it D-Branes from M-Branes}, preprint R/95/59, hep-th/
9512062. 
\item C. M. Hull and P. K. Townsend, \np 438 (1995) 109. 
\item J. Scherk and J. H. Schwarz, \np 153 (1979) 61. 
\item E. Cremmer, in {\it Supergravity 1981}, eds. S. Ferrara and J. G. Taylor
(Cambridge University Press, 1982). 
\item J. Maharana and J. H. Schwarz, \np 390 (1993) 3. 
\item J. H. Schwarz, \np 226 (1983) 269. 
\item P. Howe and P. West, \np 238 (1984) 181. 
\item C. M. Hull, \pl 357 (1995) 545. 
\item J. H. Schwarz, \pl 360 (1995) 13.
\item T. Buscher, \pl 194 (1987) 59; \pl 201 (1988) 466. 
\item M. Dine, P. Huet and N. Seiberg, \np 322 (1989) 301.
\item P. Aspinwall, {\it Some Relationship Between Dualities in String Theory},
preprint CLNS-95/1359, hep-th/9508154.
\item E. Bergshoeff, B. Janssen and T. Ortin, {\it Solution Generating
Transformations and the String Effective Action}, preprint UG-1/95, 
QMW-PH-95-1, hep-th/9506156.

\end{enumerate}

\vfil

\vfil
\eject 

\end{document}